\documentclass[prb,twocolumn,aps,showpacs]{revtex4}
\usepackage{graphicx}
\usepackage{bm}
\usepackage{amssymb} 
\begin{document}

\title{Supergap anomalies in cotunneling between 
N-S and between S-S leads via a small quantum dot}

\author{V. V. Mkhitaryan and M. E. Raikh}

\affiliation{ Department of Physics, University of Utah, Salt Lake
City, UT 84112}

\begin{abstract}
Cotunneling current through a resonant level coupled to either normal
and superconducting or to two superconducting leads is studied for the domain
of bias voltages, $V$, exceeding the superconducting gap, $2\Delta$.
Due to the on-site repulsion in the resonant level,
 cotunneling of an electron is accompanied by creation of
a quasiparticle in a superconducting lead. Energy conservation
imposes a threshold for this inelastic transport channel:
$V_c=3\Delta$ for N-S case
 and $\tilde{V}_c=4\Delta$ for the S-S case.
We demonstrate that the behavior of current near the
respective thresholds is nonanalytic, namely,
$\delta I^{in}(V)\propto
\left(V-V_c\right)^{3/2}\Theta\left(V-V_c\right)$ and
$\delta I^{in}(V) \propto (V-\tilde{V}_c)\Theta\left(V-\tilde{V}_c\right)$.
Stronger anomaly for the S-S leads is the consequence of
the enhanced density of states at the edges of the gap.
In addition, the enhanced density of states makes the
threshold anomalies for two-electron cotunneling processes
in the Coulomb-blockaded regions
more pronounced than for the N-N leads.

\end{abstract}


\maketitle

\section{Introduction}
Early single-electron-transport devices \cite{early0}
were based on
conducting grains containing large gate-controlled number
of electrons.
A grain was coupled by tunnel
barriers to two macroscopic leads.
With number of electrons on the grain being large,
superconductivity could be induced in the grain
\cite{early1,early2,early3,early4,early5,early6,early7,early8,early9}
upon lowering the temperature, while the
leads remained either normal \cite{early3,early4,early7} or also turned
into a superconducting state \cite{early1,early2,early5,early6,early8,early9}.
The focus of the early studies was
the interplay
between the two low-energy \cite{early2,early3,early4,early5,early6,early7,early8}
scales, namely, the charging energy and the
superconducting gap. This interplay manifested itself
in the Coulomb-blockade oscillations.
On the theoretical side, different regimes of
transport  via superconducting grain \cite{Averin93,matveev93,hekking93,hekking'93}
were studied for experimentally relevant situation
of a grain containing many electrons.

In the later experiments the grains have been replaced by
much smaller few-electron quantum dots, based either on
$InAs$ \cite{InAsNew1,InAsNew2,InAsNew3}
or carbon nanotubes \cite{CN1,CN2,CN3,CN4,CN5,CN6,CN7,CN8,CN9}.
In these devices, there is no superconducting pairing of
electrons on the dot. Rather, either one \cite{CN4}
or both \cite{InAsNew1,InAsNew2,InAsNew3,CN1,CN2,CN3,
CN5,CN6,CN7,CN8,CN9} leads are made of
superconducting material.

Interesting physics in
the S-N-S junctions with superconducting
leads is due to the fact \cite{0TBK} that
the Andreev process \cite{andreev,BTK} in
these junctions gives rise to a rich
subgap structure in the current-voltage
characteristics \cite{flensberg88,golub94,bratus95}.
When the N-region is a small quantum dot
(or a single resonant level) coupled by tunneling
to the leads, this subgap structure is
more pronounced \cite{aleiner96,levy97,johansson99}.
In addition, in the latter case the
on-site interaction of two
electrons, which
in a small dot assumes the role of charging energy,
becomes important
\cite{glazman89,ivanov99,rozhkov99,oganesyan02,Avishai03,kozub03,levy03,choi04,karrasch08}.


What makes the S-N-S structures with a resonant
level as a N- region
particularly interesting, is a delicate interplay  of a new energy scale,
Kondo temperature, which is much smaller than the charging energy,
and the superconducting gap. This interplay is the
focus of the very recent experimental studies
\cite{InAsNew2,InAsNew3,CN2,CN4,CN8,CN9}.
The results reported in Refs. \onlinecite{InAsNew2,InAsNew3,CN8}
suggest that subgap anomalies in differential
conductance, $G(V)$, at biases $V=\pm \Delta$,
where $2\Delta$ is the superconducting gap, are
enhanced in the Kondo regime. Another intriguing
observation made in Refs. \onlinecite{InAsNew2,InAsNew3,CN8}
is that Kondo resonance leads to smearing of the conventional
anomalies in $G(V)$ at $V=\pm 2\Delta$.

Therefore, both theoretical
\cite{glazman89,ivanov99,rozhkov99,oganesyan02,Avishai03,kozub03,levy03,choi04,karrasch08}
and experimental \cite{InAsNew2,InAsNew3,CN2,CN4,CN8,CN9} studies
suggest that  on-site
repulsion  affects the subgap structure in the
conductance. However, it is commonly believed that
for $V>2\Delta$ there is no qualitative difference
between the cases when superconducting leads are
separated by a barrier or both coupled to a quantum dot.
In the present paper we demonstrate that on-site repulsion
manifests itself even for $V>2\Delta$, leading to {\em supergap}
anomalies in $G(V)$. The underlying reason is that, at finite repulsion,
{\em inelatic} electron transitions between
normal leads become possible \cite{we}. These transitions
are accompanied by a quasiparticle excitations in the leads.
When one of the leads is superconducting, the minimal energy
of the excitation is $2\Delta$. Then, in order for electron
tunneling from the normal lead to create the excitation
in superconducting lead, the bias should exceed $V_c=3\Delta$.
Threshold for inelastic tunneling results in a supergap
singularity, $\delta G(V)\propto (V-V_c)^{1/2}$, in the N-S conductance,
as demonstrated in Sect. III A.
For the same reason, inelastic tunneling between two
superconducting leads has a threshold at $\tilde{V}_c=4\Delta$.
We show that the supergap anomaly in the S-S transport has a step-like
form, $\delta G(V)\propto\Theta(V-\tilde{V}_c)$ (Sect. IV), i.e., it is stronger
than in the N-S case.
This is due to the enhancement of the
density of states at the edges of the superconducting gap.
In addition, finite temperature, $T$, affects the S-S supergap anomaly
only via the temperature dependence of $\Delta$, whereas the
N-S supergap anomaly is a universal function of $(V-V_c)/T$ (Sect. III A.).
In Sect. III B we also demonstrate that the enhancement of the density of states
causes a sharpening of the large-bias transport anomalies \cite{we}
that involve two-electron transitions.

\begin{figure}[t]
\centerline{\includegraphics[width=48mm,angle=0,clip]{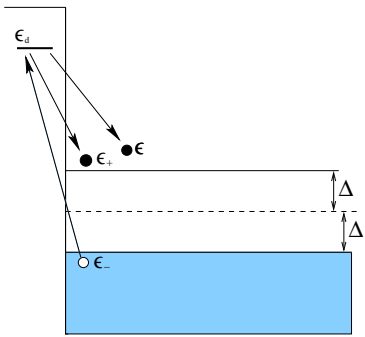}}
\caption{(Color online) Inelastic correction to the lifetime of
the localized state due to  excitation of a quasiparticle across
the gap is illustrated schematically.}
\end{figure}

\section{Anomaly in the lifetime of a localized state}

In order to illustrate how the on-site repulsion,
$U$, gives rise to the anomalies
in the conductance, $G(V)$, we start from an auxiliary
problem of the escape of an electron from the occupied
localized state (LS) into a superconductor. This situation
is illustrated in Fig. 1. If the energy of the LS, $\epsilon_d$,
lies above the upper boundary of the gap, $\epsilon_d > \Delta$,
then the population of the LS, which is occupied at time $t=0$,
decays with $t$ as
\begin{equation}\label{lifetime}
n(t)=\exp(-\Gamma t),\end{equation}
where the decay rate,
$\Gamma$, is given by the golden-rule expression
\begin{equation}
\label{Gamma}
\Gamma(\epsilon_d) =\pi \gamma^2\nu_0g(\epsilon_d).
\end{equation}
Here, $\gamma$ is the tunnel matrix element and
\begin{equation}
\label{DoS}
\nu_0g(\epsilon)=\nu_0 \frac{\epsilon}{\sqrt{\epsilon^2-\Delta^2}}
\end{equation}
is the density of states in the superconductor.
Eq.~(\ref{DoS}) applies when $\Gamma$ is much smaller
than $\Delta$. Our main point is that, for large enough $\epsilon_d$,
there exists another {\em inelastic} channel of the electron
escape into the continuum. Namely, the escape can be accompanied
by excitation of a quasiparticle across the gap. This process leads to
the
threshold anomaly in the dependence $\Gamma(\epsilon_d)$.
The position of the threshold, $\epsilon_d^{(c)}$, can be found from
the following two conditions on the energy, $\epsilon$, of electron
leaving the LS
\begin{eqnarray}
\label{conditions}
\epsilon > \Delta,~~~~~
\left(\epsilon_d-\epsilon\right)> 2\Delta.
\end{eqnarray}
The first condition ensures that the state into
which electron escapes is empty, while the meaning
of the second condition is that the energy loss
suffered by escaping electron is sufficient to
create a quasiparticle. From Eq.~(\ref{conditions})
we find the minimal value of $\epsilon_d$
\begin{equation}
\label{threshold}
\epsilon_d=\epsilon_d^{(c)}=3\Delta.
\end{equation}
Inelastic process is enabled by a finite $U$. To see
this, we notice that there are two contributions to
the amplitude of the process:

1. An electron from the LS tunnels into the state
$\epsilon>\Delta$ (i); another electron from the occupied state,
$\epsilon_-$, enters the LS (ii), and subsequently tunnels into
the empty state $\epsilon_{+}$. These steps are illustrated in
Fig.~1.

2. Initial and final states are the same as in 1, while the
intermediate steps (i) and (ii) are interchanged. As a result,
after the first step, the LS is doubly occupied. In the absence of
the on-site repulsion, the two amplitudes, 1 and 2, would cancel each
other identically. At finite $U$, this cancellation does not
happen. Note that, for large $U\gg \epsilon_d$, the energy denominator
corresponding to $\epsilon_{-}\rightarrow \epsilon_d$ contains
$U$, so that the second amplitude can be neglected.

The above reasoning is quite similar to that in
Ref.~\onlinecite{we}, where another inelastic process, occupation
of the LS in the course of cotunneling between  normal leads, has
been considered.

The amplitude, $A_{\epsilon_d, \epsilon_-}^{\epsilon,
\epsilon_+}$, of the three-step process in Fig.~1 is $\propto \gamma^3$.
Taking into account that the energies of the intermediate states
are $\epsilon$ and $\epsilon+\epsilon_d-\epsilon_-$, the
analytical expression for this amplitude reads
\begin{eqnarray}
\label{amplitude1} A_{\epsilon_d, \epsilon_-}^{\epsilon,
\epsilon_+}= \frac{\gamma^3} {(\epsilon_d-\epsilon)
(\epsilon_--\epsilon)}.
\end{eqnarray}
Note, that this expression is valid when the states, $\epsilon$
and $\epsilon_+$ correspond to the opposite spin projections
\cite{we}, so that these states are distinguishable. On the
contrary, for parallel spins of the states $\epsilon$ and
$\epsilon_+$ the amplitude Eq.~(\ref{amplitude1}) vanishes
\cite{footnote}.

The expression for inelastic correction to the rate, $\Gamma$,
follows from Eq.~(\ref{amplitude1}) 
\begin{eqnarray}
\label{rate1} \delta\Gamma(\epsilon_d)&=&
2\pi\int\limits_\Delta^\infty
d\epsilon\,\nu(\epsilon)\int\limits_\Delta^\infty
d\epsilon_+\,\nu(\epsilon_+)\int\limits^{-\Delta}_{-\infty}
d\epsilon_-\,\nu(\epsilon_-)\nonumber\\
&\times&|A_{\epsilon_d, \epsilon_-}^{\epsilon,
\epsilon_+}|^2\,\delta\,\Bigl[\epsilon_d+\epsilon_--(\epsilon+\epsilon_+)\Bigr].
\end{eqnarray}
It is seen from Eq. (\ref{rate1}) that the argument of the
$\delta$- function turns to zero for $\epsilon_d=3\Delta$ at
$\epsilon_-=-\Delta$, and $\epsilon=\epsilon_+=\Delta$. To
establish the form of the anomaly near $\epsilon_d
=\epsilon_d^{(c)}=3\Delta$, we introduce the new variables
\begin{eqnarray}
\label{newvar}E= \epsilon-\Delta,\quad E_+=
\epsilon_+-\Delta,\quad E_-= -\epsilon_--\Delta.
\end{eqnarray} in Eq. (\ref{rate1}). Now it is
sufficient to set $\epsilon_-=-\Delta$, and
$\epsilon=\epsilon_+=\Delta$ in the denominator of Eq.
(\ref{rate1}), and replace $\nu(\epsilon)$, $\nu(\epsilon_+)$, and
$\nu(\epsilon_-)$ by $\nu_0\sqrt{{\Delta}/{2E}}$,
$\nu_0\sqrt{{\Delta}/{2E_+}}$, and $\nu_0\sqrt{{\Delta}/{2E_-}}$,
respectively. Upon this replacement, Eq.~(\ref{rate1}) simplifies
to
\begin{eqnarray} \label{rate11} \delta\Gamma(\epsilon_d)&=&
\frac{\Gamma^3}{2^{\,9/2}\pi^2\Delta^{5/2}}\int\limits_0^\infty
\frac{dE}{\sqrt{E}}\int\limits_0^\infty
\frac{dE_+}{\sqrt{E_+}}\int\limits_{0}^{\infty}
\frac{dE_-}{\sqrt{E_-}}\nonumber\\
&\times&\delta\,\Bigl[\epsilon_d-\epsilon_d^{(c)}-\bigl(E+E_++
E_-\bigr)\Bigr].
\end{eqnarray} The above integral is proportional to
$\left(\epsilon_d-\epsilon_d^{(c)}\right)^{1/2}$; the numerical
factor can be easily expressed through the surface area of the
unit sphere. The final form of the threshold anomaly is the
following
\begin{eqnarray}
\label{sqrt} \frac{\delta\Gamma(\epsilon_d)}{\Gamma}=
\frac{\Gamma^2}{2^{\,7/2}\pi\Delta^{5/2}}
\left[\,\epsilon_d-\epsilon_d^{(c)}\right]^{1/2}\Theta\left(
\epsilon_d-\epsilon_d^{(c)}\right).
\end{eqnarray} In deriving Eq.~(\ref{sqrt}) we assumed that the
intrinsic width, $\Gamma$, is much smaller than $\Delta$. This
guarantees that the relative correction $\delta\Gamma/\Gamma$ is
small. The anomaly Eq. (\ref{sqrt}) is much stronger than the
threshold anomaly for two-electron ionization of the LS in Ref.
\onlinecite{we}. The origin of this enhancement is the divergence
of the density of states Eq.~(\ref{DoS}) at edges of the gap.

In the above calculation we treated the states $\epsilon_-$,
$\epsilon_+$, and $\epsilon$ as electron states in a normal metal,
and took superconductivity into account only via the energy
dependence of the density of states, $\nu(\epsilon)$. This is
justified when the tunneling amplitude is calculated to the lowest
order in the matrix element, $\gamma$. However the anomaly
Eq.~(\ref{sqrt}) emerges in the third order in $\gamma$. The proof
of the validity of Eq.~(\ref{rate1}) for
$\delta\Gamma(\epsilon_d)$, starting from the BCS Hamiltonian, is
presented in the Appendix.

\section{ Supergap anomalies in the N-S cotunneling}
\subsection{Single-electron transport}

Passage of current from a metal to a superconductor by
single-electron transitions, involving the LS, is illustrated in
Fig. 2.
Position, $V_c$, of the anomaly, at which the cotunneling from
the normal lead can be accompanied by creation of a quasiparticle
in the superconducting lead, can be found from the similar reasoning as in
Sect.~I. The only difference is that electron enters the
superconducting lead with energy close the Fermi energy of the
normal lead, so that
\begin{equation}
\label{CONDITION}
V_c=3\Delta.
\end{equation}
The magnitude of the anomaly
is, however, weaker than for the electron escape considered in
Sect. I. This is due to the fact that, while the energy of the LS
is fixed to $\epsilon_d$,  the energy of the electron in the normal lead
is simply restricted to the domain
below $V/2$ - the Fermi level in the normal lead.

\begin{figure}[t]
\centerline{\includegraphics[width=50mm,angle=0,clip]{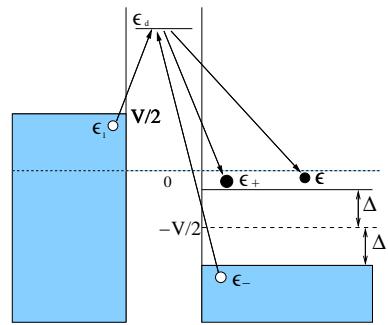}}
\caption{(Color online) Origin of the anomaly at $V_c=3\Delta$ in
cotunneling between the N and S leads is illustrated
schematically.}
\end{figure}

The elastic cotunneling conductance is given by
\begin{eqnarray}
\label{elcon} G^{el}_{NS}=\frac{4e^2}{\pi\hbar}
\frac{\Gamma_L\Gamma^s_R}{(\epsilon_d-V/2)^2},
\end{eqnarray} where we assumed $(\epsilon_d-V/2)\ll V$.
The widths $\Gamma_{L,R}=\pi\nu_{L,R} \gamma^2_{L,R}$ are defined
in a usual way; due to the enhancement of the density of states in
the superconductor the width, $\Gamma^s_{R}$, which enters into
Eq. (\ref{elcon}), becomes $\Gamma^s_{R}=\pi\nu_Rg(V)
\gamma^2_{R}\approx 3\Gamma_R/\sqrt{8}$.

In order to calculate the inelastic correction, $\delta
G^{in}(V)$, to the conductance, one cannot simply modify
$\Gamma^s_{R}$ according to Eq.~(\ref{sqrt}). This is because, in
the course of cotunneling, the electron occupies the LS only {\em
virtually}. The correct procedure of finding $\delta G^{in}(V)$
requires calculation of  inelastic correction, $\delta
I^{in}(V)$, to the current,  taking into account that electron,
transferred from the normal into superconducting lead, can excite a
quasiparticle in this lead. Then we have
\begin{eqnarray}
\label{incur} \delta I^{in}(V)\!\!&=& \frac {4\pi
e}{\hbar}~\nu_L\nu_R^3\hspace{-.2cm}\int\limits_{-\infty}^{\infty}
\hspace{-.1cm}d\epsilon_1f(\epsilon_1-V/2)\hspace{-.4cm}
\int\limits_{\Delta-V/2}^{\infty}\hspace{-.3cm}d\epsilon~
g(\epsilon +V/2)\nonumber\\
&\times&\hspace{-.5cm}\int\limits_{\Delta-V/2}^{\infty}\hspace{-.2cm}d
\epsilon_+g(\epsilon_++V/2)\hspace{-.3cm}
\int\limits^{-\Delta-V/2}_{-\infty} \hspace{-.2cm}d\epsilon_-g(\epsilon_-+V/2)\nonumber\\
&\times&|A_{\epsilon_1,\epsilon_-}^{\epsilon,\epsilon_+}|^2
\delta\Bigl[\epsilon_1+\epsilon_--(\epsilon+\epsilon_+)\Bigr],
\end{eqnarray} where $f(\epsilon)$ is the Fermi function.
The expression for the transition amplitude $(\epsilon_1,\epsilon_-)\rightarrow
(\epsilon,\epsilon_+)$ differs from Eq.~(\ref{amplitude1}) by an extra
$\gamma_L$, namely
\begin{eqnarray}
\label{amplitude2} A_{\epsilon_1, \epsilon_-}^{\epsilon,
\epsilon_+}=&& \frac{\gamma_L\gamma_R^3} {(\epsilon_d-\epsilon_1)
(\epsilon-\epsilon_1)(\epsilon_d-\epsilon_+)}\\
&&+\frac{\gamma_L\gamma_R^3} {(\epsilon_d-\epsilon_-)
(\epsilon_+-\epsilon_-)(\epsilon_d-\epsilon)}.\nonumber
\end{eqnarray}
As in Sect. I, in Eq. (\ref{amplitude2})
we had excluded the virtual states with doubly occupied LS.
Two terms in Eq.~(\ref{amplitude2}) account for two different sequences
in which the transition $(\epsilon_1,\epsilon_-)\rightarrow
(\epsilon,\epsilon_+)$  takes place. The first term corresponds to
electron from the normal lead entering the LS at the first step.
The second term describes virtual occupation of the LS by
electron from superconductor with energy $\epsilon_{-}$ at the first step,
followed by its escape into $\epsilon_{+}$ and subsequent cotunneling
of electron from the normal lead. Note, that there is no analog of the
second contribution in the amplitude Eq.~(\ref{amplitude1}). This is
because Eq.~(\ref{amplitude1}) describes the
process in which the LS was occupied in the initial state.
\begin{figure}[t]
\centerline{\includegraphics[width=60mm,angle=0,clip]{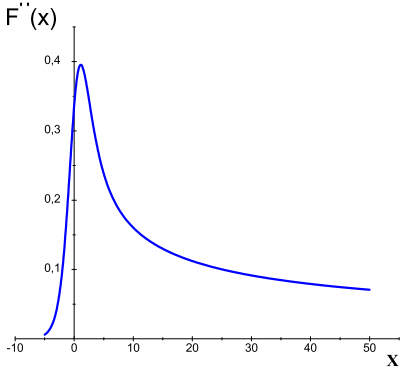}}
\caption{ (Color online) The shape of the peak in the derivative,
$ {d G}/{d V}$, of the N-S differential conductance is plotted
from Eq.~(\ref{tlfl}) versus dimensionless deviation $x=(V-V_c)/T$
}
\end{figure}

In order to extract the anomaly, upon substituting Eq.~(\ref{amplitude2})
into Eq.~(\ref{incur}), we introduce the new variables
\begin{eqnarray}
\label{newvar}E_1= \epsilon_1-V/2,\quad E= \epsilon + V/2
-\Delta,\\ E_+= \epsilon_+ + V/2-\Delta,\quad E_-= -\epsilon_--
V/2-\Delta.\nonumber
\end{eqnarray} For bias, $V$, close to $V_c=3\Delta$,
characteristic values of $E$, $E_1$, $E_+$, and $E_-$ are much
smaller than $\Delta$. This allows to set $\epsilon_1=V_c/2$ and
$\epsilon=\epsilon_+=-V_c/2+\Delta$ in the denominators of Eq.~(\ref{amplitude2}).
We can also use the near-gap-edge asymptotes
for the densities of states in the superconducting leads. After
these simplifications, Eq.~(\ref{incur}) assumes the form
\begin{eqnarray}
\label{incur1}&&\!\!\!\delta I^{in}(V)=\frac{2^{\,5/2}}
{\pi^3}\,\frac e{\hbar}\\
&&\times\frac {\Gamma_L\Gamma_R^3\, (\Delta T)^{3/2}}
{\bigl[(\epsilon_d-3\Delta/2)
\bigl((\epsilon_d+3\Delta/2)^2-\Delta^2\bigr)\bigr]^2}\,
\text{\Large F}\!\left(\frac{V-V_c}T\right)\!,\nonumber
\end{eqnarray}
where the dimensionless function $\text{\large F}$ of a single
argument, $(V-V_c)/T$, is defined as
\begin{eqnarray}\label{tlf}
\text{\Large F}\left(\frac{V-V_c}T\right)\!\!\!
&=&\!\!\!\!\int\limits_{-\infty}^{\infty}
\frac{dE_1f(E_1)}{T^{3/2}} \int\limits_0^\infty\!\!
\frac{dE}{\sqrt{E}}\int\limits_0^\infty\!\!
\frac{dE_+}{\sqrt{E_+}}\int\limits_{0}^{\infty}\!\!
\frac{dE_-}{\sqrt{E_-}}\nonumber\\
\times&& \hspace{-.5cm}\delta\,\Bigl[V-V_c+E_1-
\bigl(E+E_++E_-\bigr)\Bigr].
\end{eqnarray} Note, that the three-fold integration over $E$,
$E_+$, and $E_-$ has already been carried out in Sect. 1. It
yields $2\pi(V-V_c-E_1)^{1/2} \Theta\,\Bigl[V-V_c+E_1\Bigr]$. As a
result, the bias dependence of $\delta I^{in}$ is
given by a single integral
\begin{eqnarray}\label{tlfl}
\text{\Large F}\left(\frac{V-V_c}T\right)=2\pi \int\limits_{0}^{\infty}
dx\frac {\sqrt{x}}{\exp\bigl[x-\frac{V-V_c}T\bigr]+1}.\quad
\end{eqnarray}
Inelastic correction , $\delta G^{in}(V)$, to the differential
conductance is thus described by the derivative, $d\text{\large F}/d
V$. The asymptotic behavior of $\delta G^{in}(V)$ at low,
$(V-V_c)\ll T$, and high,  $(V-V_c)\gg T$, temperatures can be
easily found from Eq.~(\ref{tlfl}). We present the results for a
dimensionless ratio, $\delta G^{in}/G^{el}$, of inelastic and
elastic contributions to the conductance
\begin{eqnarray} \label{condasympt}&&\hspace{-.6cm}
\frac{\delta G^{in}}{G^{el}}= \left\{
\begin{array}{l}
\!\!\!\frac {2\Delta^2\Gamma_R^2} {3\pi\bigl[(\epsilon_d+
3\Delta/2)^2-\Delta^2 \bigr]^2}
\sqrt{\frac{V-V_c}\Delta},\quad (V-V_c)\gg T,\nonumber\\
\,\\
\hspace{-.12cm}\frac
{2\Delta^2\Gamma_R^2} {3\pi^{1/2}\bigl[(\epsilon_d+ 3\Delta/2)^2-\Delta^2
\bigr]^2}\sqrt{\frac T\Delta},\quad (V-V_c)\ll T, \nonumber
\end{array}
\right.\\&&\;
\end{eqnarray} where $\alpha=2^{-1/2}\int_0^\infty dx\, x^{1/2}/\cosh^2x\approx 0.536$.
It is seen from Eq.~(\ref{condasympt}) that, at $V>V_c$,
differential conductance acquires a correction $\propto
(V-V_c)^{1/2}$. Correspondingly, the second derivative, $d^2 I/d
V^2$, has an asymmetric peak of a width $\sim T$ centered at
$V=V_c$. The shape of the peak is given by the second derivative
of the function ${\large F}$. In Fig.~3 this derivative,
calculated numerically from Eq.~(\ref{tlfl}), is plotted versus
dimensionless deviation, $(V-V_c)/T$.

\begin{figure}[t]
\centerline{\includegraphics[width=70mm,angle=0,clip]{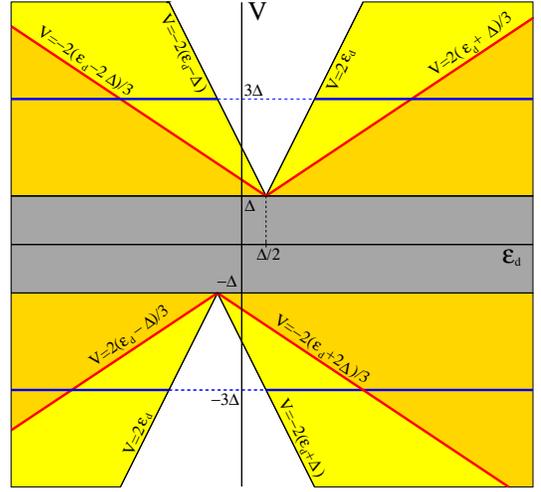}}
\caption{(Color online) Stability diagram for transport  between N
and S leads via a localized state. White region corresponds to the
sequential tunneling transport. Horizontal blue lines, $V=\pm
3\Delta$, correspond to the {\it single-electron} supergap
anomaly, illustrated in Fig.~2. Red lines, $V=2/3(\epsilon_d\pm
\Delta)$ and $V=2/3(\epsilon_d\pm 2\Delta)$, are the positions of
the {\it two-electron} resonance. Subgap resonances at $|V|<
\Delta$ lie in the shaded region.}
\end{figure}

\subsection{Two-electron transport}
\subsubsection{Ionization of the LS}
In terms of the Coulomb blockade stability diagram in the
$(\epsilon_d,\, V)$ plane, Fig.~4, the anomalies at $V=\pm 3\Delta$
correspond to horizontal lines, which start from the points
$(-\Delta/2,\,3\Delta)$, $(3\Delta/2,\,3\Delta)$, and
$(-3\Delta/2,\,-3\Delta)$, $(\Delta/2,\,-3\Delta)$. These lines
extend into the blockaded region. In Ref.~\onlinecite{we} it was
demonstrated that, without superconductivity, there exists an
additional weak structure within the Coulomb blockade diamond,
along the lines $V= \pm 2\epsilon_d/3$. The origin of this
structure is the {\it two-electron} ionization of the LS, namely,
the process, in which one electron from the left lead is
transferred to the right lead while the other electron from the
left lead occupies the LS. The position of the boundary, $V=
2\epsilon_d/3$, expresses the threshold for this two-electron
transfer, which follows from the energy conservation.
In this subsection we point out that, in the presence
of the superconductivity, the boundaries for two-electron
ionization are modified in an {\it asymmetric} fashion. For
positive bias, $V>0$, the boundaries are located at
\begin{equation}\label{+modbound} V_{+}(\epsilon_d)=\pm
 \frac23\left(\epsilon_d-\frac\Delta2\right)+\Delta,
\end{equation} while for negative bias they are located at
\begin{equation}\label{-modbound}
V_{-}(\epsilon_d)=\mp \frac23\left(\epsilon_d+\frac\Delta2\right)
-\Delta.
\end{equation} These modified boundaries are shown in Fig.~4. More
importantly, as we demonstrate below, superconductivity leads to
the {\it strengthening} of the ionization anomaly. The underlying
mechanism for this strengthening is, again, the enhancement of
the density of states at the boundaries of the gap.

Energy dependence of the density of states can be easily
incorporated into the expression from Ref.~\onlinecite{we} for
ionization rate. Consider first the situation when the initial
states of two electrons with energies $\epsilon_1$ and
$\epsilon_2$ are in the normal lead, while one of the finite
states (with energy $\epsilon$) is in the superconducting lead and
the other is on the LS. The ionization rate for $T=0$ is given by
\begin{eqnarray}
\label{NSir} \Gamma_{ion}^{{\scriptscriptstyle N\rightarrow
S}}(V)= \frac {\Gamma_L^2\Gamma_R}
{(2\pi)^2}\hspace{-.1cm}\int\limits_{-\infty}^{V/2}
\hspace{-.1cm}d\epsilon_1 \int\limits_{-\infty}^{V/2}
\hspace{-.1cm}d\epsilon_2 \hspace{-.3cm}
\int\limits_{\Delta-V/2}^{\infty}\hspace{-.3cm}d\epsilon~
g(\epsilon +V/2)\nonumber\\
\times\frac1{(\epsilon_d-\epsilon_1)^2(\epsilon_d-\epsilon_2)^2}\,
\delta\Bigl[\epsilon_d+\epsilon -\epsilon_1-\epsilon_2\Bigr].
\end{eqnarray}
Near the threshold, $V=V_{+}(\epsilon_d)$, one can set
$\epsilon_1=\epsilon_2=V_{+}/2$ in the denominator of
Eq.~(\ref{NSir}). Upon measuring the energies $\epsilon_1$,
$\epsilon_2$, and $\epsilon$ from their respective boundaries,
as in Eq.~(\ref{newvar}), we can simplify Eq.~(\ref{NSir}) to
\begin{eqnarray}
\label{NSir1} \Gamma_{ion}^{{\scriptscriptstyle N\rightarrow
S}}(V)=\frac
{\Gamma_L^2\Gamma_R\Delta^2}{(2\pi)^2(\epsilon_d-V_+/2)^4}
\text{\Large H}_{+}\left[\frac{V-V_+}\Delta \right],
\end{eqnarray}
where the one-parameter function, $\text{\large H}_{+}$, is
defined as
\begin{eqnarray}
\label{H+} \text{\large H}_{+} (x)
=\!\!\int\limits_0^{\infty}\!\frac {dE_1}\Delta\!
\int\limits_0^{\infty}\!\frac {dE_2}\Delta g\left( \frac32\Delta\,
x+\Delta-E_1-E_2 \right)
\nonumber\\
\times \Theta\!\left[ \frac32\,x-\frac{E_1}\Delta - \frac{E_2}
\Delta \right].
\end{eqnarray} This integral is easily calculable. Its analytic
form is \begin{eqnarray}\label{anH+} \text{\large H}_{+}(x)&=&
\frac{3x+2}8\sqrt{9x^2+12x}\\
&&-\frac12\ln\left[\frac32 x+1+\sqrt{\frac94
x^2+3x}\,\right].\nonumber
\end{eqnarray} The large- $x$ and small- $x$ asymptotes of $\text{\large H}_{+}$ are
\begin{eqnarray} \label{H+asympt}
\text{\large H}_{+}(x)= \left\{
\begin{array}{l}
\sqrt{3}\, x^{3/2},\quad x\ll 1,\\
\,\\
\frac98\, x^{2},\quad x\gg 1,
\end{array}
\right.
\end{eqnarray}

\begin{figure}[t]
\centerline{\includegraphics[width=60mm,angle=0,clip]{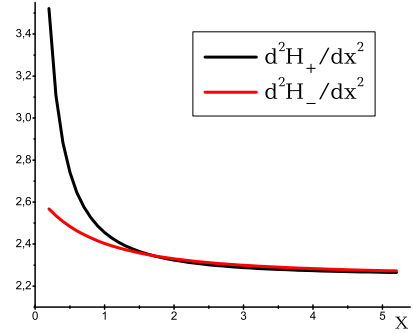}}
\caption{(Color online) Shapes of the anomalies in $d^2I/dV^2$
near $V=V_{+}$ and $V=V_{-}$ versus dimensionless deviations
$x=(V-V_{\pm})/\Delta$ calculated, respectively, from Eqs.
(\ref{anH+}) (black line), and (\ref{FinH-}) (red line). Both curves approach the
value $9/4$ at $x\rightarrow \infty$. For small $x$, $d^2H_{+}/dx^2$ diverges as
$3^{3/2}/4x^{1/2}$, while the $x=0$ value of $d^2H_{-}/dx^2$ is $27\pi/32$. Note that
$H_{+}(x)$ and $H_{-}(x)$ are zero for $x<0$.}
\end{figure}

Consider now $V<0$. The ionization rate is given by the expression
\begin{eqnarray}
\label{SN} &&\Gamma_{ion}^{{\scriptscriptstyle S\rightarrow
N}}(V)= \frac {\Gamma_L\Gamma_R^2}
{(2\pi)^2}\hspace{-.1cm}\int\limits_{-\infty}^{-V/2-\Delta}
\hspace{-.4cm}d\epsilon_1\, g(\epsilon_1 +V/2) \hspace{-.4cm}
\nonumber\\
&&\times\int\limits_{-\infty}^{-V/2-\Delta}
\hspace{-.4cm}d\epsilon_2\, g(\epsilon_2 +V/2)
\int\limits_{V/2}^{\infty}\hspace{-.1cm}d\epsilon~\frac
{\delta\Bigl(\epsilon_d+\epsilon -\epsilon_1-\epsilon_2\Bigr)}
{(\epsilon_d-\epsilon_1)^2(\epsilon_d-\epsilon_2)^2}, \nonumber\\
\end{eqnarray}
which differs from Eq.~(\ref {NSir}) by additional density of
superconducting states  in the integrand. When the bias voltage is
near the critical, $V=V_-$, one can replace the values of $\epsilon_1$ and
$\epsilon_2$ by their boundary value $-V_-/2- \Delta$, in the denominator of
Eq.~(\ref{SN}). Then the ionization rate, $\Gamma_{ion}^{{\scriptscriptstyle S\rightarrow
N}}$, can be expressed as
\begin{eqnarray}
\label{NSir2} \Gamma_{ion}^{{\scriptscriptstyle S\rightarrow
N}}(V)=\frac
{\Gamma_L\Gamma_R^2\Delta^2}{(2\pi)^2(\epsilon_d+V_-/2 +\Delta)^4}
\text{\Large H}_{-}\left[\frac{V_--V}\Delta \right],\nonumber\\
\end{eqnarray} where we have absorbed all the integrals of Eq.~(\ref
{SN}) into the new one-parameter function, $\text{\large H}_{-}$,
defined as follows: \begin{eqnarray} \label{H-} &&\hspace{-1cm}\text{\large
H}_{-}(x)=\!\!\int\limits_0^{\infty}\!\frac {dE_1}\Delta
\,g(E_1+\Delta)\! \int\limits_0^{\infty}\!\frac {dE_2}\Delta
\,g(E_2+\Delta)
\nonumber\\
&&\times \Theta\!\left[ \frac32\,x-\frac{E_1}\Delta -
\frac{E_2}\Delta \right].
\end{eqnarray} One of the integrations in Eq.~(\ref{H-}) can be
performed explicitly. The final form of the function $\text{\large
H}_{-}$ is the following \begin{eqnarray}
\label{FinH-}
\text{\large H}_{-}\! (x) =\!\!\int\limits_0^{3x/2}\!\! {dz}\frac
{(z+1)\sqrt{3x/2-z}\,\sqrt{3x/2+2-z}}{\sqrt{z^2+2z}}.
\end{eqnarray} The easiest way to find the behavior of $\text{\large
H}_{-}(x)$ at large and small $x$ is to set, respectively,
$g(\epsilon)=1$ and $g(\epsilon)=\sqrt{\Delta/2\epsilon}$ in the
integrand of the definition Eq.~(\ref{H-}). This yields
\begin{eqnarray} \label{H-asympt}
\text{\large H}_-(x)= \left\{
\begin{array}{l}
\frac{3\pi}4x\left (1+\frac{9}{16}x\right),\quad x\ll 1,\\
\,\\
\frac98\, x^{2},\quad x\gg 1,
\end{array}
\right.
\end{eqnarray} Upon populating the LS, the electron rapidly,
within the time $(\Gamma_L+\Gamma_R)^{-1}$, escapes either to the
left or to the right lead. In terms of contributions to inelastic
current these two channels of escape are different\cite{we}.
For escape to the left, the net charge transfer is $e$,
while for escape to the right, it is $2e$.
As a result, the inelastic contribution to the
current is equal to $\delta I^{in}(V)= 2e\Gamma_{ion}(V)(2\Gamma_R
+\Gamma_L)/(\Gamma_L+\Gamma_R)$. The threshold behavior of $\delta
I^{in}$ near $V_+$ and $V_-$ is determined by the functions $H_+$
and $H_-$, respectively. As seen from Eqs.~(\ref{H+asympt}) and
(\ref{H-asympt}), these behaviors coincide when $V-V_+$, $V-V_-$
are much bigger than $\Delta$. This is natural since for large
deviations from the thresholds, superconducting gap drops out from
$\delta I^{in}$. Note, however, that in the immediate vicinities
of $V_+$ and $V_-$, the threshold behaviors are {\it different},
namely, $\delta I^{in}$ is more singular near $V_-$ than near
$V_+$. The origin of this asymmetry is that the inelastic process
$N\rightarrow S$ involves only one state near the superconducting
gap, while the inelastic process $S\rightarrow N$ involves two
such states. Without superconductivity, threshold anomaly at
$V=\pm 2\epsilon_d/3$ shows up in the third derivative of the
current with respect to $V$. Our results, Eqs.~(\ref{NSir}) and
(\ref{NSir2}), suggest that, within the interval $\sim \Delta$
from the thresholds $V_+$, $V_-$, the singularities of current are
more pronounced: they show up already in the second derivative
$d^2I/d V^2$. This is illustrated in Fig. 5, where
$d^2\text{\large H}_{+}/d V^2$ and $d^2\text{\large H}_{-}/d V^2$
are plotted.

In fact, the singular behavior of inelastic current near $V=V_{-}$
shows up already on the level of differential conductance, $\delta G^{in}= d\delta I^{in}/dV$,
as a step $\propto \Theta(V_{-}-V)$.
Combining Eq.~(\ref{NSir2})   and Eq.~(\ref{H-asympt}), we get the following magnitude of the step

\begin{eqnarray}
\label{SNfs}\delta G^{in}(V)=\frac{3e^2}{16\pi\hbar}\frac
{\Gamma_L\Gamma_R^2\Delta}{(\epsilon_d+V_-/2 +\Delta)^4} \Theta(V_{-}-V).
\end{eqnarray}

\subsubsection{Two-electron tunneling}

As seen from Eqs.~(\ref{NSir1}) and (\ref{H+}), the relative correction, $\delta
I^{in}(V)/I^{el}$, to the elastic current due to ionization of the LS,
changes on the scale $V\sim\Delta$; the magnitude of correction at
$V\sim\Delta$ being $\sim \Gamma_R\Delta/\epsilon_d^2\ll1$.
Although small, this correction is distinguishable by virtue of
its threshold dependence on bias. Indeed, both $\text{\large
H}_{+}$ and $\text{\large H}_{-}$ are zero for $V<V_+$ and
$V>V_-$, respectively. Another fact that distinguishes 
the transport at biases near $V_{+}$ and $V_{-}$ is that the
inelastic current, $\delta I^{in}(V)$,  has a precursor
with singular dependence on deviation $V-V_{\pm}$ and on 
the temperature $T$. The origin of this precursor\cite{we} is
direct cotunneling of {\em two} electrons via the LS.
This process differs from ionization of the LS, since,
in course of this two-electron  cotunneling, the LS is 
populated only virtually. As a result, the corresponding
contribution to the current, $I^{2e}_{\pm}(V)$, contains
extra power $\Gamma_L$ (or $\Gamma_R$). On the other hand,
this contribution is more singular in deviation, $V-V_{\pm}$,
and has a peculiar $T$-dependence. As all other corrections
to the elastic cotunneling calculated above, $I^{2e}_{\pm}(V)$
is enabled by a finite on-site repulsion.
The golden-rule expression for  $I^{2e}_{+}(V)$
\begin{eqnarray}
\label{NStun} &&\delta I^{2e}_+(V)= \frac e {\hbar}\frac
{\Gamma_L^2\Gamma_R(2\Gamma_R +\Gamma_L)}
{(2\pi)^2(\epsilon_d-V_+/2)^2(V_++\Delta)^2}
\nonumber\\
&&\hspace{-.4cm}\times
\hspace{-.1cm}\int\limits_{-\infty}^{\infty}
\hspace{-.1cm}d\epsilon_1\,f(\epsilon_1-V/2)\hspace{-.2cm}
\int\limits_{-\infty}^{\infty}
\hspace{-.1cm}d\epsilon_2\,f(\epsilon_2-V/2)\\
&&\hspace{-.4cm}\times \hspace{-.5cm}
\int\limits_{-V/2+\Delta}^{\infty}\hspace{-.4cm}dE_1\, g(E_1
+V/2)\hspace{-.2cm}\int\limits_{-\infty}^{\infty}\hspace{-.1cm}d
E_2 \frac{\delta\Bigl(\epsilon_1+\epsilon_2 -E_1-E_2\Bigr)}
{(\epsilon_d-E_2)^2}\nonumber
\end{eqnarray}
contains energy denominators that correspond to virtual states; in these states
the LS is occupied by first and then by second tunneling
electron. Note, that the $\delta$-function in Eq.~(\ref{NStun}) ensures
conservation of the {\em total} energy, $\epsilon_1+\epsilon_2$, of
two electrons in the initial and final states, while individual energies
get redistributed. Sensitivity of  $\delta I^{2e}_+(V)$ to $V=V_{+}$ comes
from the domain of integration in Eq.~(\ref{NStun}) with $E_1$ near the Fermi edge, 
$E_1\approx -V_{+}/2+\Delta$, and $E_2\approx \epsilon_d$. For this reason, the
nonresonant energy denominators are extracted from the integrand of Eq.~(\ref{NStun}).

In order to capture the dependence of $\delta I^{2e}_+$ on $(V-V_+)$ and $T$, 
we introduce the dimensionless function, $\text{\large K}_{+}$, defined as
\begin{eqnarray}
\label{K+}\text{\large K}_{+}(x)= \int\limits_{-\infty}^{\infty}
\frac{dz_1}{e^{z_1}+1}\int\limits_{-\infty}^{\infty}
\frac{dz_2}{e^{z_2}+1}\int\limits_{0}^{\infty}\frac{dz_3}{\sqrt{z_3}}
\int\limits^{\infty}_{-\infty} \frac{dz_4}{z_4^2}\nonumber\\
\times \delta\Bigl[\frac32x +z_1+z_2-z_3-z_4\Bigr].
\end{eqnarray} Then $\delta I^{2e}_+(V)$ can be presented in the
form \begin{eqnarray}\label{I-K} &&\hspace{-0.2cm}\delta
I^{2e}_+(V)= \frac e {\hbar}\frac {\Gamma_L^2\Gamma_R(2\Gamma_R
+\Gamma_L)}
{(2\pi)^2(\epsilon_d-V_+/2)^2(V_++\Delta)^2}\left(\frac{\Delta
T}2\right)^{1/2}\nonumber\\
&&\hspace{2cm} \times\,\,\text{\Large K}_{+}
\left[\frac{V-V_+}{T}\right].\end{eqnarray} 
As before, in Eq.~(\ref{I-K}) we used the near-gap-edge asymptote 
of $g(\epsilon)$, so that  Eq.~(\ref{I-K}) applies in the interval
$\Gamma_L, \Gamma_R \ll \vert V-V_+\vert, T \ll \Delta$. 

The four-fold
integration in Eq.~(\ref{K+}) can be reduces to a single integral by
using the Fourier representation for the $\delta$- function and
the fact that the Fourier transform of the Fermi function is equal
to $\tilde{f}(\omega)=\pi/ {\sinh(\pi\omega)}$. We will present the
result for the second derivative, $d^2\text{\large K}_{+}/dx^2$, 
which describes the near-threshold behavior of  $d^2I^{2e}_{+}/d V^2$. It reads
\begin{eqnarray}
\label{ddK+}\frac {d^2\text{\large K}_{+}}{dx^2}\!\!\!\!\!\!&&=
-\frac{9 \pi^{5/2}}{2^{5/2}}\\ &&\times \int\limits_{0}^{\infty}ds
\frac{s^{5/2}} {\sinh^2(\pi
s)}\left[\cos\left(\frac32sx\right)+\sin\left(\frac32sx\right)
\right].\nonumber
\end{eqnarray}
\begin{figure}[t]
\centerline{\includegraphics[width=60mm,angle=0,clip]{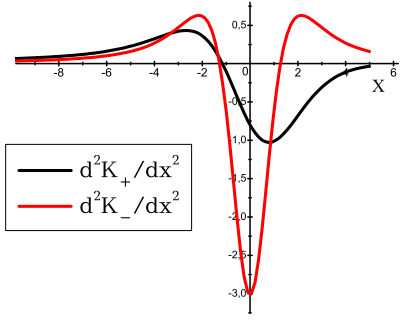}}
\caption{(Color online) Shapes of the anomalies in
$d^2I^{2e}_{\pm}/dV^2$ near $V=V_{+}$ and $V=V_{-}$ versus
dimensionless deviations $x=(V-V_{\pm})/T$ calculated,
respectively, from Eqs. (\ref{K+}) (black line), and (\ref {K-})
(red line).}
\end{figure}
Consider first the limiting case of vanishing $T$. 
To realize that the temperature drops out from the
expression Eq.~(\ref{I-K}), we notice that the asymptotic
behavior of $d^2\text{\large K}_{+}/dx^2$  at large 
{\em negative} $x$ is $\propto \vert x\vert^{-3/2}$.  
This yields $d^2\delta I^{2e}_+/dV^2\propto (V-V_+)^{-3/2}$.
The divergence is stronger than $1/(V-V_+)$ in Ref.~\onlinecite{we}.
Another remarkable feature of $d^2I^{2e}_{+}/dV^2$ is that,
at finite $T$, it exhibits a fine structure. This is seen from
 Fig. 6, where the function $d^2\text{\large K}_{+}/dx^2$ is plotted.
Asymptotic behavior of $d^2\text{\large K}_{+}/dx^2$ at large positive $x$
is $\propto \exp(-3x/2)$. This suggests that for $V>V_+$ ionization current
dominates over $\delta I^{2e}_+$.

Calculation of the two-electron current, $\delta I^{2e}_-(V)$,
near $V=V_{-}$ is quite similar to Eqs. (\ref{NStun}) and (\ref{I-K})   
Namely, the golden-rule expression
\begin{widetext} 
\begin{eqnarray}
\label{SNtun} \delta I^{2e}_-(V)\!\!\!&=&\!\!\! \frac e {\hbar}\frac
{\Gamma_L\Gamma_R^2(\Gamma_R +2\Gamma_L)}
{(2\pi)^2(\epsilon_d+V_-/2+\Delta)^2(V_-+\Delta)^2}\hspace{-.3cm}
\int\limits_{-\infty}^{-V/2-\Delta}
\hspace{-.3cm}d\epsilon_1\,g(\epsilon_1+V/2)\hspace{-.3cm}\int\limits_{-\infty}^{-V/2-\Delta}
\hspace{-.3cm}d\epsilon_2\,g(\epsilon_2+V/2)\nonumber\\
&&\times\hspace{-.2cm}\int\limits_{-\infty}^{\infty}\hspace{-.1cm}dE_1\,
\bigl[1-f(E_1-V/2)\bigr]\hspace{-.2cm}
\int\limits_{-\infty}^{\infty}\hspace{-.1cm}d E_2
\frac{\delta\Bigl(\epsilon_1+\epsilon_2 -E_1-E_2\Bigr)}
{(\epsilon_d-E_2)^2}
\end{eqnarray}
\end{widetext}
is cast into the form
\begin{eqnarray}\label{I-K} \delta I^{2e}_-(V)= \frac e
{\hbar}\frac {\Gamma_L\Gamma_R^2(\Gamma_R +2\Gamma_L)\Delta}
{8\pi^2(\epsilon_d+V_-/2+\Delta)^2(V_-+\Delta)^2}\nonumber\\
\times\,\text{\Large K}_{-}
\left[\frac{V_--V}{T}\right].\end{eqnarray}
The dimensionless function $\text{\large K}_{-}(x)$ is
a four-fold integral over the electron energies (in the units of $T$)
in the initial and final states 
\begin{eqnarray}
\label{K-}\text{\large K}_{-}(x)= \int\limits_{0}^{\infty}
\frac{dz_1}{\sqrt{z_1}}\int\limits_{0}^{\infty}
\frac{dz_2}{\sqrt{z_2}}
\int\limits_{-\infty}^{\infty}\frac{dz_3}{1+e^{-z_3}}
\int\limits^{\infty}_{-\infty} \frac{dz_4}{z_4^2}\nonumber\\
\times\, \delta\,\Bigl[\frac32x +z_1+z_2+z_3+z_4\Bigr].
\end{eqnarray} Three out of four integrations in Eq.~(\ref{K-}), over $z_1$, $z_2$,
and $z_4$, can be carried out explicitly. Then we get
\begin{eqnarray}
\label{ddK-}\frac {d^2\text{\large K}_{-}}{dx^2}=
-\frac{9\pi^2}{4}\int\limits_{0}^{\infty}ds \frac{s^2} {\sinh(\pi
s)}\cos\left(\frac32sx\right).
\end{eqnarray}
 Second derivative, $d^2 \text{\large K}_{-}/dx^2$, is plotted
in Fig. 6. It shows that $d^2I^{2e}_{-}/dV^2$ also exhibits 
a fine structure at $(V_--V) \sim T$. The low-$T$ behavior of
$d^2I^{2e}_{-}/dV^2$ is even more singular than that of 
$d^2I^{2e}_{+}/dV^2$. This follows from the large-$x$ asymptote, 
$\propto 1/x^2$, of the integral Eq.~(\ref{ddK-}).  Thus, the
below-threshold behavior of $d^2I^{2e}_{-}/dV^2$ is $\propto 1/(V_--V)^2$.

\section{Anomaly in the S-S cotunneling}

\begin{figure}[t]
\centerline{\includegraphics[width=60mm,angle=0,clip]{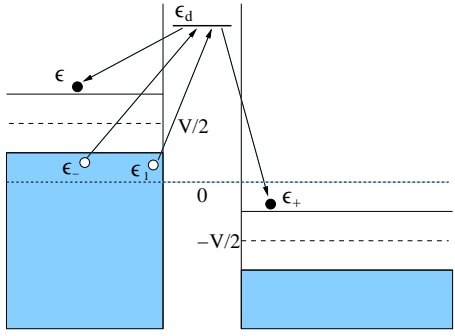}}
\caption{(Color online) One of the possible inelastic channels in
the S-S transport is illustrated schematically. Electron transfer
is accompanied by creation of the excitation in the left lead.}
\end{figure}


Energy diagram for transport between two superconducting leads via an LS
is shown in Fig.~7 for bias $V>2\Delta$. Similarly to the
case of normal and superconducting leads, electron cotunneling can be accompanied
by excitation of a quasiparticle across the gap. It is easy to see
from Fig. 7 that the threshold bias for this process is
\begin{eqnarray}
\label{SSvc} \tilde{V}_c=4\Delta. 
\end{eqnarray} 
The difference from the N-S case is
that, at the threshold, electron tunnels from the edge of the gap
rather than from the Fermi level of the metal. A more significant
qualitative difference from the N-S geometry is that a quasiparticle
can be excited in {\it both} leads. Besides, as we will see below,
the anomaly is stronger in the S-S than in the N-S case. This
is due to the divergence of the density of states in both leads.
More specifically, instead of the four-fold integral
Eq.~(\ref{incur}), the near-threshold expression for inelastic contribution to
the current reads
\begin{widetext}
\begin{eqnarray}
\label{ssincur} \delta I^{in}(V)\!\!\!&=&\!\!\!\frac e {\hbar}
\frac{4\,\Gamma_R^3\Gamma_L}{\pi^3\bigl[(\epsilon_d+3\Delta)
(\epsilon_d^2-\Delta^2)\bigr]^2}\hspace{-.3cm}
\int\limits_{-\infty}^{V/2-\Delta}
\hspace{-.3cm}d\epsilon_1g(\epsilon_1-V/2)\int\limits_{\Delta-V/2}^{\infty}\hspace{-.3cm}d\epsilon~
g(\epsilon +V/2)\hspace{-.3cm}\int\limits_{\Delta-V/2}^{\infty}
\hspace{-.2cm}d\epsilon_+g(\epsilon_++V/2)\nonumber\\
&&\times\int\limits^{-\Delta-V/2}_{-\infty}
\hspace{-.4cm}d\epsilon_-g(\epsilon_-+V/2)\,
\delta\Bigl[\epsilon_1+\epsilon_-
-(\epsilon+\epsilon_+)\Bigr].
\end{eqnarray} 
\end{widetext}
 Similarly to Eq.~(\ref{incur}), in order to
calculate the integral Eq.~(\ref{ssincur}), we introduce the same
variables $E$, $E_+$, $E_-$ as in Eq.~(\ref{newvar}), and also
$\tilde{E}_1= -(\epsilon_1-V/2+\Delta)$. Upon taking the near-gap
asymptotes for the density of states, Eq.~(\ref{ssincur}) assumes
the form
\begin{eqnarray} \label{4E}
&&\delta I^{in}(V)=\frac e {\hbar}
\frac{\Gamma_R^3\Gamma_L\Delta^2}{\pi^3\bigl[(\epsilon_d+3\Delta)
(\epsilon_d^2-\Delta^2)\bigr]^2}\!\! \int\limits_{0}^{\infty}\!\!
\frac{d\tilde{E}_1}{\sqrt{\tilde{E}_1}} \int\limits_0^\infty\!\!
\frac{dE}{\sqrt{E}}\nonumber\\
&&\times\!\!\int\limits_0^\infty
\!\!\frac{dE_+}{\sqrt{E_+}}\!\int\limits_{0}^{\infty}
\!\!\frac{dE_-}{\sqrt{E_-}}\,\delta\Bigl[V-\tilde{V}_c-
\bigl(\tilde{E}_1+E+E_++E_-\bigr)\Bigr].\nonumber\\
\end{eqnarray} After rescaling all variables to $(V- \tilde{V}_c)$, this
integral reduces to the surface area of a unit sphere in four
dimensions, and we obtain: \begin{eqnarray} \label{ssfin} \delta
I^{in}(V)=\frac e {\hbar}
\frac{\Gamma_R^3\Gamma_L\Delta^2}{12\pi^2\bigl[(\epsilon_d+3\Delta)
(\epsilon_d^2-\Delta^2)\bigr]^2}\nonumber\\ \times(V-\tilde{V}_c)
\Theta\,\Bigl[V-\tilde{V}_c\Bigr].
\end{eqnarray} Contribution Eq.~(\ref{ssfin}) describes cotunneling accompanied by excitation 
of a quasiparticle in the right lead. Similar calculation for inelastic channel, with
excitation of a quasiparticle, as depicted in Fig.~7, 
results in
\begin{eqnarray} \label{ssfinF} \delta
I^{in}(V)=\frac e {\hbar}
\frac{\Gamma_R\Gamma_L^3\Delta^2}{12\pi^2\bigl[(\epsilon_d-3\Delta)
(\epsilon_d^2-\Delta^2)\bigr]^2}\nonumber\\ \times(V-\tilde{V}_c)
\Theta\,\Bigl[V-\tilde{V}_c\Bigr].
\end{eqnarray} Here we would like to emphasize that both calculations leading to 
Eqs.~(\ref{ssfin}) and (\ref{ssfinF}) take into account that quasiparticle can be created at the 
first as well as at the last step of the cotunneling process, and corresponding {\it amplitudes 
interfere}, as in Eq.~(\ref{amplitude2}). Taking this interference into account, results in the 
extra factor $\sim \Delta^2/\epsilon_d^2$ in Eqs.~(\ref{ssfin}) and (\ref{ssfinF}).
Obviously, the threshold anomaly Eq.~(\ref{ssfin}) in the current
results in the jump in the $V$-dependence of the differential
conductance. Within a numerical factor and assuming
$\epsilon_d\gg\Delta$, the magnitude of the jump can be presented
as \begin{eqnarray} \label{magnitude} \left( \frac{\delta
G^{in}}{G^{el}}\Bigg|_{\tilde{V}_c^+} -\frac{\delta
G^{in}}{G^{el}}\Bigg|_{\tilde{V}_c^-} \right) \sim\,\,\,\frac
{(\Gamma_L^2+\Gamma_R^2)\Delta^2}{\epsilon^4_d}.\end{eqnarray} 
Here the sum
$\Gamma_L^2+\Gamma_R^2$ accounts for the contributions of the two
channels of inelastic current, mentioned above. We note, that the
step Eq.~(\ref{magnitude}) is abrupt; its temperature smearing is
$\propto \exp(-\Delta/T)$ rather than $\sim T$, as in the case of
tunneling between N and S leads.

Overall, the stability diagram for superconducting leads 
differs from Fig.~4 in two respects. Firstly, the positions
of the supergap anomalies are $\tilde{V}_c=\pm 4\Delta$.
Secondly, the stability diagram is {\em symmetric} with respect
to $V\rightarrow -V$. Namely, the boundaries of the two-electron
ionization anomaly in this case are located at 
$V_c^{\ast}=\pm \frac{2}{3}\left(\epsilon_d+3\Delta\right)$.   
Regarding the ``strength'' of two-electron anomaly, the threshold behavior of the 
differential conductance can be found from calculation similar to Eqs.~(\ref{NSir}),
(\ref{SN}), and  within a prefactor yields
\begin{equation}
\label{withoutC}
\delta G^{in}(V) \propto \left(V-V_c^{\ast}\right)^{-1/2},
\end{equation} 
i.e., the threshold behavior is
more singular than Eq.~(\ref{SNfs}). Again, the divergence of $\delta G^{in}$
is limited by $\left(V-V_c^{\ast}\right) \sim \Gamma_{L,R}$ rather than by temperature.    

\section{Concluding remarks}

Let us list the assumptions adopted in the above consideration:

\noindent(i) energy position, $\epsilon_d$, of the LS is well outside the
superconducting gap, $\Delta$;

\noindent(ii) on-site repulsion, $U$, is the largest energy scale,
$U\gg|\epsilon_d|$;

\noindent(iii) the widths, $\Gamma_L$, $\Gamma_R$, are the smallest energy
scales, so that
\begin{equation}\label{3scales} \Gamma_L, \Gamma_R\ll\Delta\ll
\epsilon_d\ll U.\end{equation} One of the consequences of
Eq.~(\ref{3scales}) is that the Kondo temperature, $T_K\propto
\exp[-\pi|\epsilon_d|/2(\Gamma_L+ \Gamma_R)]$, is much smaller
than $\Delta$. This means that the Kondo effect will not developed
fully, but rather manifest itself as an enhancement $\propto
\ln^{-2}\Delta/T_K$ of the conductance at small bias.

Recent experimental papers Refs.
\onlinecite{InAsNew2,InAsNew3,CN2,CN4,CN8,CN9} are focused on the
domain of parameters $T_K\sim\Delta$, where the two prominent
regimes of transport compete with each other.
This competition is due to the fact that antiparrallel
spins of electrons in the Cooper pairs cannot mediate the
spin-flip processes that are responsible for the Kondo effect.
Experimentally, in the case of normal leads, the Kondo effect
manifests itself on the stability diagram in the $(\epsilon_d, V)$
plane as enhanced zero-bias conductance in the valley
$\epsilon_d<0$, where LS is occupied. It has  no effect on the
valley $\epsilon_d>0$. On the other hand, with superconducting
leads, conductance is suppressed in the entire domain of biases
$V<2\Delta$ in both valleys. A non-trivial result of interplay
between the Kondo effect and superconductivity is that
the peaks at $V=\pm\Delta$ emerge in the Kondo valleys,
whereas the conventional peaks at $V=\pm 2\Delta$ are
suppressed \cite{InAsNew2,InAsNew3,CN8}.
This implies that the Andreev transport process is
facilitated by the Kondo resonance. Conversely,
in the non-Kondo valleys, the peaks $V=\pm\Delta$
do not show up, while $V=\pm 2\Delta$-peaks are strong
and exhibit a well-known threshold behavior, reflecting
the BSC density of states.

In the present paper we predict additional anomalies {\em both} outside
the Kondo regime and above the gap. Nevertheless,
the origin of the new anomalies is intimately
related to the Kondo physics.
To clarify this relation, we recall
that, in a {\em bulk} metal with
magnetic impurities the
energy exchange between electrons
is possible even without direct
electron-electron interaction.
This was first demonstrated by Kaminski
and Glazman in Ref.~\onlinecite{kaminski01}.
Obviously, such an exchange is impossible in the case
of non-magnetic impurities. The reason is that
the mechanism, which is responsible for an impurity (LS)
being magnetic, is a finite on-site repulsion, $U$.
As a result, the interaction between two electrons
in metal, leading to the energy exchange, takes
place when they {\em virtually} visit the LS.
The energy exchange occurs between electrons
with opposite spins, and in the case of magnetic
impurity, involves spin-flips~\cite{kaminski01}.
Thus the mechanism Ref.~\onlinecite{kaminski01}
represents the most elementary manifestation of
the Kondo physics, and even does not require
the presence of the Fermi sea.

As was demonstrated in Ref.~\onlinecite{we},
the mechanism~\cite{kaminski01} can be extended
to the transport between two normal leads, coupled
to the LS. Then, for two electrons tunneling
between the leads,  the magnitude of the energy exchange
is limited by the applied bias, $V$. This leads to
the anomalies in conductance at $V=\pm 2\epsilon_d/3$.
The main message of the present paper is that, in the
case when one or both leads are superconducting, the
gap, $2\Delta$, sets the threshold for inelastic process
of one-electron transfer accompanied by a quasiparticle
excitation in the superconducting lead. The ensuing
anomalies at $V_c=\pm 3\Delta$ (for N and S leads) and
at $\tilde{V}_c=\pm 4\Delta$ (for S-S leads) are independent of
the gate voltage, $\epsilon_d$.
The anomaly  near $V=\tilde{V}_c$ is not smeared by temperature
and manifests itself as a sharp peak in the second derivative
$d^2I(V)/dV^2$. Although the papers on transport through
Coulomb-blockaded dots report the data on first derivative, i.e., the
differential conductance, $dI(V)/dV$, the second derivative
was previously measured for single-electron transport
through a molecule \cite{SM}. In Ref. \onlinecite{SM} the second
derivative was required to resolve a fine structure in the $I(V)$-dependence,
related to the vibrational satellites.

As a final remark, we note that higher-order, in parameters, $\Gamma_L/\Delta$,
$\Gamma_R/\Delta$ processes will lead to anomalies at even larger
biases due to creation of more than one quasiparticle by a tunneling
electron.  For the case of the S-S leads, additional
anomalies can be expected at biases $V_c^{(n)}=2\Delta(2 + n)$.
Estimate for the behavior of inelastic current $(V-V_c^{(n)})\ll \Delta$
can be easily found by extending the  four-fold integral in Eq. (\ref{4E})
to higher $n$. This yields:
$\delta I_n^{in}(V) \propto (V-V_c^{(n)})^{n+1}\Theta(V-V_c^n)$.

\acknowledgments We gratefully  acknowledge 
 useful discussions with E.~G.~ Mishchenko and F.~von~Oppen.


\section{Appendix}

The fact that the superconductivity manifests itself in the
expression Eq.~(\ref{Gamma}) for the tunneling rate
$\Gamma(\epsilon_d)$ {\it only} through the density of states
Eq.~(\ref{DoS}) is well known. However, it is not obvious that
higher-order, in the tunnel matrix element, $\gamma$, corrections
to $\Gamma(\epsilon_d)$ can be expressed solely through
$g(\epsilon)$, and do not contain coherence factors. Indeed, in
our calculations we treated the amplitude, $A_{\epsilon_d,
\epsilon_-}^{\epsilon, \epsilon_+}$, Eq.~(\ref{amplitude1})  as a
number determined only by the energies
\begin{equation}\label{disp}
\epsilon\,(\xi)=\pm\sqrt{\Delta^2+\xi^2}
\end{equation}
of initial and final states {\it in superconductor}, and ignored
the fact that the real amplitude contains contributions of
positive and negative bare energies $\xi$. This contributions
enter into the amplitude with different weights, namely
\begin{equation}
\label{cfu} u(\xi)=\frac1{\sqrt{2}}\left[1+
\frac\xi{\sqrt{\Delta^2+\xi^2}} \right]^{1/2},
\end{equation} for the upper branch in Eq.~(\ref{disp}) and
\begin{equation}
\label{cfv} v(\xi)=\frac1{\sqrt{2}}\left[1-
\frac\xi{\sqrt{\Delta^2+\xi^2}} \right]^{1/2},
\end{equation} for the lower branch in Eq.~(\ref{disp}). Then, when performing
summation over states corresponding to, say, upper branch, one
has to take into account contributions $\propto u(\xi)$ and
$\propto u(-\xi)$, since they correspond to the same energy
$\epsilon=\sqrt{\Delta^2+\xi^2}$. Now the fact that the main
contribution $\Gamma(\epsilon_d)$ Eq.~(\ref{Gamma}) to the
lifetime does not contain coherence factors can be formally
interpreted as a consequence of the identity $u^2(\xi)+u^2(-\xi)
=1$.

Turning to the third-order amplitude Eq.~(\ref{amplitude1}),
the correct way to write one particular contribution to
$A_{\epsilon_d, \epsilon_-}^{\epsilon, \epsilon_+}$ is
\begin{eqnarray}
\label{SCamp}&& A_{\epsilon_d, \epsilon_-}^{\epsilon,
\epsilon_+}(\xi_-,\,\xi,\,\xi_+) = \\
&&\frac{\gamma^3\,v(\xi_-)\,u(\xi)\, u(\xi_+)}
{\left(\epsilon_d-\sqrt{\Delta^2+\xi^2}\,\right) \left(-
\sqrt{\Delta^2+\xi_-^2}-\sqrt{\Delta^2+\xi^2}\,\right)}. \nonumber
\end{eqnarray} Then the correction $\delta\Gamma(\epsilon_d)$ is,
actually, the sum of all possible contributions, i.e.,
\begin{eqnarray}
\label{corrlt}&&\delta\Gamma(\epsilon_d)\propto
\sum\limits_{{\xi_->0}\atop{\xi_-<0}}
\sum\limits_{{\xi>0}\atop{\xi<0}}
\sum\limits_{{\xi_+>0}\atop{\xi_+<0}}\big | A_{\epsilon_d,
\epsilon_-}^{\epsilon, \epsilon_+}(\xi_-,\,\xi,\,\xi_+)\big |^2.
\end{eqnarray} From Eq.~(\ref{corrlt}) it becomes apparent that
coherence factors in the numerators of eight contributions can be
combined into the product $[u^2(\xi_-)+u^2(-\xi_-)]
[u^2(\xi)+u^2(-\xi)] [u^2(\xi_+)+u^2(-\xi_+)]$, which is an
identical unity. Note, that this conclusion rests on the
assumption that the matrix element, $\gamma$, is independent of
$\xi$.


\begin{references}
\bibitem{footnote} Strictly speaking, this reasoning applies for the
case of tunneling into a metal. In fact, for tunneling into a
superconductor, one has an electron from the LS with definite spin
projection and a broken Cooper pair in the spin-singlet state.
\end{references}
\end{document}